\documentclass[a4paper,10pt]{article}
\usepackage{graphicx} 
\usepackage{graphics}

\usepackage{amsmath}   
\usepackage{amsfonts}
\usepackage{amssymb}
 
\begin{document}

\setlength{\oddsidemargin} {1cm}
\setlength{\textwidth}{18cm}
\setlength{\textheight}{23cm}
\title{Scaling Regimes as obtained from the DR5 Sloan Digital Sky Survey.}

\author{{\bf Reuben Thieberger$^{a,1}$ and Marie-No\"elle C\'el\'erier$^{b,2}$} \\ {\small $^a$ Physics Department, Ben Gurion University, Beer Sheva, 84105 Israel} \\
{\small $^b$ Laboratoire Univers et TH\'eories (LUTH) Observatoire de Paris,} \\
{\small 5 place Jules Janssen, 92190 Meudon, France}\\
{\small e-mail: $^1$ thieb@bgumail.bgu.ac.il} \\
{\small $^2$ marie-noelle.celerier@obspm.fr}}

\maketitle

\begin{abstract}

Standard cosmology is based on the assumption that the universe is spatially homogeneous. However the consensus on a homogeneous matter structure, even on very large scales, has never been complete. The advantage of correlation dimension calculations is that they enable one to obtain the transition scale to a homogeneous distribution, whereas other methods, such as those using the two-point correlation function, make it hard to exhibit the possible fractal properties of the Universe matter content. Our purpose is to calculate the correlation dimension $D_2$, looking for a possible transition to homogeneity, which would imply $D_2 = 3$. We apply the correlation integral method to the three dimensional sample composed of 332,876 galaxies which we extract from the Fifth Data Release of the Sloan Digital Sky Survey. We analyze the raw data up to the scale $d_{max} = 160$ Mpc, assuming $H_0 = 70$ km s$^{-1}$ Mpc$^{-1}$ and considering three cosmological models in order to test the model dependence of our method. Using volume limited samples for this range leaves us with about 20,000 galaxies. Applying our method to random maps helps us to calibrate our results. We obtain a correlation dimension of the galaxy distribution which seems to increase with scales up to $D_2=3$ reached around 70 Mpc. The results of our analysis, performed on the largest volume limited sample which can be extracted from the SDSS catalog, are compatible with those formerly obtained by other authors. However, to get a more reliable description of the structures at various scales, we think it will be mandatory to use still larger catalogs than those which are currently available.


\end{abstract}

\section{Introduction} \label{i}

Standard cosmology is based on the assumption that the Universe is 
spatially homogeneous, at least on scales sufficiently large to 
justify its approximation by a Friedmann-Lema\^itre-Robertson-Walker 
(FLRW) model. The high \linebreak isotropy measured in the cosmic microwave 
background radiation (CMBR) is usually considered as a strong evidence 
in support of this hypothesis.

The structures seen in galaxy catalogs - groups, clusters and 
super-clusters, distributed along voids, filaments and walls -
are not viewed as contradicting this principle, as the common
opinion is that the scales on which the universe is assumed
to be homogeneous are much larger than those subtended by 
these structures. However, the consensus on a 
homogeneous feature of matter, even on very large scales, 
has never been complete (see, e.g., Pietronero, 1987). At least, 
the value of the transition scale from inhomogeneity to homogeneity 
needs to be tested (C\'el\'erier, 2000; Romano, 2007).

One of our motivations for the present study comes from our
interest in trying to distinguish the different scaling regimes
that may exist in the galaxy distribution, though on different
length scales. Our interest was first raised by the apparent 
discrepancy in the collection of results obtained 
from the analyses of data realized by a number of authors, 
which appeared to differ essentially, not only by the statistical 
methods employed, but mostly by the scales spanned by the studied samples.

A little more than twenty years ago, three-dimensional galaxy catalogs have been 
made available. They supported the first findings of approximate 
self-similarity in the large-scale distribution of galaxies
within prescribed scale intervals (Provenzale, 1991).

It was claimed, from an analysis of the data of the 
CfA redshift survey catalog, worked out with the correlation function 
method, that, at small separations, i.e., for a correlation length 
$r_0=5h^{-1}$ Mpc, the galaxy distribution exhibits a fractal structure 
with dimension $D_2\sim 1.2$  (e.g., Davis and Peebles, 1983). Other methods applied to the same  catalog, and, in particular, the correlation 
integral method (Coleman, Pietronero and Sanders, 1988) gave a fractal 
dimension slightly larger, $D_2 \sim 1.3$ to $1.5$. Using a set of 
measurements from the Observatory of Nice, a value of $D_2 \sim 2$ was 
found (Thieberger, Spiegel and Smith, 1990). Pietronero and his collaborators (Sylos Labini, Montuori and Pietronero, 1998) proceeded then to the analysis of all the currently available redshift surveys and claimed that the galaxy distribution 
exhibits a constant correlation dimension $D_2 \sim 2$ up to scales of at least 
$150h^{-1}$Mpc. These results have been widely discussed in review 
articles published in 1999 (Martinez, 1999; Wu, Lahav and Rees, 1999).

It was also suggested that three scaling regimes may be discerned in the galaxy distribution. For the Perseus-Pisces redshift survey and a reanalysis of other published results, the values ran from  $D_2 \sim 1.2$ at small separations, i. e., up to $3.5 h^{-1}$ Mpc, to $D_2 \sim 2.2$ on larger scales, with homogeneity seemingly reached at scales $>30 h^{-1}$ Mpc (Guzzo et al., 1991). Murante et al. (1998) claimed that on the smallest scales the results are consistent with a distribution of density singularities, that in the intermediate range there is a scaling behavior suggestive of flat structures such as Zeldovitch (1970) favored and that on the largest scales, the data seem to indicate a homogeneous galaxy distribution.

The last few years have seen a dramatic increase in the number and depth of galaxies with known redshifts. Therefore, many authors studied the galaxy clustering, using the matter power spectrum as an indicator (see, e. g., Tegmark et al., 2002; Tegmark et al., 2004; Sylos Labini et al., 2007; Percival et al., 2007). However, this method is not adapted to compute accurately correlation dimensions (Pietronero, 1987). Therefore, other authors calculated the correlation dimension $D_2$ for different scales using various methods (see Jones et al., 2005, for a review). A comparison of all these results shows that this dimension is unambiguously scale-dependent and increases from values less than 2 at scales less than 10 Mpc (Martinez and Jones, 1990; Guzzo et al. 1991) to values approaching 3 at much larger scales (Pan and Coles, 2000; Hogg et al., 2005; Tikhonov, 2006).

However, it has to be admitted that these data may as yet not be adequate to clearly decide such an issue and that is why we wish to reexamine it in the light of one of the currently available wide catalogs. We use here the DR5 sample from the Sloan Digital Sky Survey (SDSS, 2006) to enable us to repeat older calculations and hope to obtain more reliable results.

To test the influence of cosmological distortion, already discussed from a theoretical point of view in (Spedalere and Schucking, 1980; Ribeiro, 1995; C\'el\'erier and Thieberger, 2001), we perform our calculations for three cosmological models, the Euclidean Universe, The Einstein-de Sitter model and the $\Lambda$CDM model, and we compare the results. We also study the other possible bias or systematic errors.

We devote the next Sec.~\ref{cim} to a short reminder of the correlation integral 
method. In Sec.~\ref{ts} we describe the sample we extract from the catalog to complete our study. The possible bias and systematic errors are discussed in Sec.~\ref{bse}, before describing our analysis of the sample and giving our main results in Sec.~\ref{ags}. We test these results in Sec.~\ref{armc} by comparing them to the analysis of a mock catalog completed with the same method. In Sec.~\ref{dc}, we give a discussion and our conclusion.

\section{The Correlation Integral Method.} \label{cim}

The advantage of correlation dimension calculations is that they enable one
to obtain the transition scale to a homogeneous distribution, whereas the usual
method (two-point correlation function calculation) makes it hard to exhibit the possible fractal properties of the Universe (Pietronero, 1987).

Thieberger, Spiegel and Smith (1990) discussed the different approaches to calculate correlation dimensions. The two methods most commonly used are
the correlation integral (Grassberger and Procaccia, 1983) and the
$\Gamma$ method used originally by Pietronero (Pietronero, 1987).
That paper gives a comparison between the various methods
using well-known fractal sets for which we are able to derive
the exact dimension. We choose here to use, as a characterization of the 
of point set structures, the {\it correlation integral} 
(Grassberger and Procaccia, 1983), defined as:

\begin{equation}
C_2(r)= {1\over {N^{\prime}(N-1)}}\sum_i \sum_{j\ne i}
\Theta(r - |{\bf X}_i - {\bf X}_j|),
\label{cor}
\end{equation}
where $\Theta$ is the Heaviside function. The inner summation is over 
the whole set of $N-1$ galaxies with coordinates ${\bf X}_j$, 
$j\ne i$, and the outer summation is over a subset of $N^{\prime}$ 
galaxies, taken as centers, with coordinates ${\bf X}_i$. By 
taking only the inner $N^{\prime}$ galaxies as centers we allow 
for the effect of sample finiteness, i. e., 
for the exclusion of the points close to the edges (see, e. g., Provenzale 
et al., 1997; Sylos Labini, Montuori and Pietronero, 1998). 
As it is mentioned in those references, there are other methods for taking the edges into account. They
enable one to consider larger volumes, but one pays out by not knowing exactly
what are the approximations involved in it. We will return to discuss
these points in the data section.
In  Eq.~(\ref{cor}) $r$ goes from a small value to $r_{max}$. Since $r_{max}$ is not too large (typically 700 Mpc), light cone effects might be small enough to be ignored (C\'el\'erier and Thieberger, 2001). However, we will check this assumption by testing three different cosmological models in our subsequent calculations.

We would like to stress that this characterization is also valid when
the set is not fractal. Therefore it seems to us appropriate
to use this approach in all cases of analyzing galaxies as point sets.

Now, we may interpret $C_2(r)$ as ${\cal N}(r)/N$ where ${\cal N}(r)$ is the
average number of galaxies within a distance $r$ of a typical galaxy in
the set.  As $r$ goes to zero, $C_2$ should also vanish and, for general
distributions, we express this property as $C_2 \propto r^{D_2}$. For
computational purposes it is more convenient to use the form:

\begin{equation}
\log (C_2)=CONST. +D_2 \log (r).
\label{lsq}
\end{equation}

The exponent $D_2$ is called the correlation dimension
and it is necessarily $\le 3$ for an embedding space of dimension three.
When $D_2$ is not equal to the topological dimension 3 of space 
in our Universe, the distribution is called fractal and $D_2$ is 
its correlation dimension (Mandelbrot, 1982). 

In Eq.~(\ref{cor}) we have a double summation. In some special cases, e. g., when one has a pencil beam catalog, the inner region has to be very small and, in the worst case, it can happen that one has to use the
distances from a mere single point. This of course results in a strong
deterioration of the statistics. Fortunately, the SDSS catalogue is wide 
enough and we are not confronted with this problem here.

\section{The Sample Extracted from the SDSS Catalog} \label{ts}

The data we analyze in this paper are extracted from the fifth release of the Sloan Digital Sky Survey (SDSS, 2006).

The Sloan Digital Sky Survey (York et al. 2000) is an ongoing imaging survey of approximately $\pi$ steradians of the sky in five band pass: $ugriz$ (Fukugita et al. 1996; Gunn et al. 1998; Smith et al. 2002). Its aim is also to obtain spectroscopic spectra (Richards et al. 2002; Blanton et al. 2003) of about one million of the objects detected by photometric monitoring (Hogg et al. 2001). The rms galaxy redshift errors are $\sim$ 30 km ${\rm s}^{-1}$ and hence negligible for our purpose. The images are processed by automated pipelines which yield photometric calibrations (Lupton et al. 2001; Smith et al. 2002; Stoughton et al. 2002; Pier et al. 2003; Ivezi\'c et al. 2004) allowing one to select the galaxies from images of stars and quasars (Eisenstein et al. 2001; Strauss et al. 2002). For each object, the position, several flux varieties, morphological parameters and a provisional classification are given. The catalog also includes informational flags on each pixel and object (Stoughton et al. 2002).

The main galaxy sample (Strauss et al. 2002) focuses on galaxies with apparent magnitude brighter than $m = 17.77$ for which spectroscopic spectra are available. The sample we use for our study is extracted from this main galaxy sample, which contains 465,789 galaxies, from which we make the following eliminations:

$ 0.02 \leq z \leq 0.22$ , $ m \leq 17.77$

zWarning = $ zW=0$,

zStatus = $zS>2$,

zConfidence = $zC>0.96$.

The primary flux measure used for galaxies in this catalog is the SDSS Petrosian magnitude, petroMag, which is a modified version of the quantity defined by Petrosian (1976). When atmospheric seeing is null, the Petrosian magnitude gives a constant fraction of a galaxy light regardless of distance or size (Blanton et al. 2001; Strauss et al. 2002). Since galaxies bright enough to be included in the SDSS spectroscopic sample have relatively high signal-to-noise ratio measurements of their Petrosian magnitude up to $m=$ 20 or so (Stoughton et al. 2002), we choose to perform the data analysis reported below with the following determination of the apparent magnitude of each galaxy in our sample: $m \equiv$ petroMag - extinction (atmospheric absorption).

The zWarning flag includes several empirical tests to determine if the provided galaxy redshift is reasonable. Of course, choosing zWarning = 0 cuts several objects that may be of interest (low signal-to-noise objects for example). However, it is a conservative choice which allows us to keep in the sample only objects with accurate redshift determination.

The zStatus flag indicates what is the current status of the redshift determination for each object. Above zS = 2, the redshift is available in a consistent manner.

The zConfidence flag is a measure of the statistical confidence level in the redshift measurements.

A review of the SDSS imaging strategy and of the composition and properties of the main galaxy sample can be found in Tegmark et al. (2004) \footnote {Please note the difference between our notation and Tegmark et al.'s who call $r$ the apparent magnitude we denote as $m$.}. See also (SDSS, 2006) for a more thorough description of the catalog and an explanation of the above notations.

After completing these eliminations, we are left with 332,876 galaxies which comprise the catalog with which we work.

\section{Bias and Systematic Errors} \label{bse}

\subsection{Redshift-space Distortions}

Gravitation, which makes galaxies cluster, also causes them to move with respect to the Hubble flow. Their so-called peculiar velocities thus yield an anisotropy of their clustering in redshift space. Even if this effect implies that nonlinear corrections cannot be neglected at the rather small scales where structures virialize, it can be exactly modeled and accounted for on the large scales on which clustering is linear and which are of interest for us (Kaiser 1987).

This effect has been thoroughly studied by Tegmark et al. (2004), using a sample of SDSS galaxies. They have shown that, although estimates of the redshift-space distortions are very sensitive to nonlinear effects, an estimate of the real-space matter power is not. They have checked that any scale-dependent statistical bias in their real-space matter power spectrum due to nonlinear redshift distortions is smaller than a few percent for the scales under study and that the systematic errors associated with this effect are negligible in comparison with the statistical errors. We therefore rely on this result to ignore this effect in our analysis.

Moreover, for large scales, where the linear Kaiser effect is at work, we have tested it on our results by considering three cosmological models with different distance-redshift relations. We show thus that these results are model-independent (see Sec.~\ref{ags}) and therefore that we can neglect this effect in our analyses.

\subsection{Galaxy Bias}

The main aim of this work is to calculate the correlation dimension(s) of the galaxy distribution in view of possibly determining the transition from inhomogeneity to homogeneity which should appear at scales of order (some?) 100 Mpc.

A more detailed analysis of the overall clustering amplitude is also potentially interesting, e. g., to constrain cosmological models, but it is beyond the scope of our study. This normalization is known to be a strong function of both galaxy color and luminosity (Norberg et al.2001, 2002; Zehavi et al. 2002). Therefore, an understanding of galaxy bias is needed to derive cosmological constraints from this statistic.

On the smallest scales, where non linear corrections to the matter power spectrum are large, the shapes of galaxy power spectra are known to depend on galaxy color and luminosity (e. g., Cole et al. 2005). On larger scales, which are of interest for us, these effects are more uncertain. We therefore choose to ignore galaxy bias in our calculations.

\subsection{Systematic Errors in the Data}

Now, we must contemplate the possibility that systematic errors occur in the data. Such effects include radial modulations of the density field due to misestimates of evolution or K-corrections and angular modulations due to miscorrected dust extinction, variable observing conditions, photometric calibration errors or fiber collisions.

They have been analyzed by Tegmark et al. (2004) who showed that their effect on the estimate of the matter power spectrum is negligible. We therefore ignore these systematics in our study. 

Moreover, since we are only interested in large scale behavior, the problems whose effects are limited to small scales, such as fiber collisions, that reduce the survey ability to measure redshifts for very close pairs (below 55 arcsec), are irrelevant in our case.

\subsection{Sample Incompleteness}

The parameters available in the catalog to locate each object are the right ascension, RA, the declination, DEC, and the redshift, $z$.

However, another source for possible inaccuracies in the results of our analysis is sample incompleteness. To deal with this issue, and select the most possible complete sample among the previously selected data, we proceed as follows.

\begin{figure}
\centering
\includegraphics[height=6cm,width=8cm]{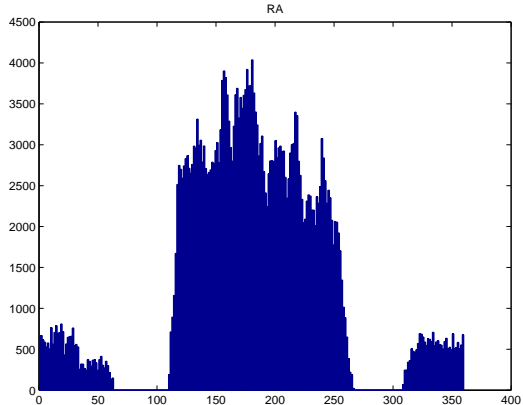}
\caption{The histogram of RA.  }
\label{Fig1}
\end{figure}

First, we map the histograms of these three variables. Figure $\ref{Fig1}$ shows, as an example, the histogram of RA.

After examining the three histograms, we choose:

$115<RA<250$

$25<DEC<60$

$0.04<z<0.16$

so that the remaining sample looks the most complete in this three-dimensional space.

\section{Analysis of the Galaxy Sample} \label{ags}

Here we assume $H_0$ = 70 km s$^{-1}$ Mpc$^{-1}$. Note that the exact value retained for $H_0$ is not too important since another value would only have as an effect to shift the transition scale the ratio of both values.

We have already noticed that the fact that the minimum of $z$ is sufficiently large
makes the problem of redshift distortion for close galaxies irrelevant.  Since C\'el\'erier and Thieberger (2001) demonstrated that for a distance, $r_{max}$, up to at least over 600 Mpc., relativistic corrections to the Euclidean distance  approximation are negligible for the calculations performed here, we might also expect that at large z, which are also in our case quite limited, redshift distortion should not be too significant. However, we want to check this point and, for this purpose, we perform our calculations for three different cosmological models. We then compare the results to see if we find a significant difference between the three results, i.e., much larger than our statistical errors.

The three models we put to the test are: the Euclidean, Einstein-de Sitter and $\Lambda$CDM models. Their luminosity distances as a function of redshift are:

\textbf{The Euclidean (or redshift) distance}
\begin{equation}
D_E=\frac{cz}{H_0}
\label{euclid}
\end{equation}

\textbf{The Einstein-de Sitter luminosity distance} (C\'el\'erier and Thieberger, 2001)
\begin{equation}
D_{EdS}=\frac{2c}{H_0}(1+z-\sqrt{1+z})
\label{eds}
\end{equation}

\textbf{The $\Lambda$CDM luminosity distance} (C\'el\'erier, 2000)
\begin{eqnarray}
D_{\Lambda CDM}= \frac{c}{H_0}z + \frac{c}{4H_0}(2 - \Omega_M + 
2\Omega_{\Lambda})z^2 + \frac{c}{8H_0}F_3(\Omega_M,\Omega_{\Lambda})\nonumber \\
F_3(X,Y)=(-2X -4Y -4XY + X^2 + 4Y^2)z^3 +{\cal O}(z^4)
\label{Lcdm}
\end{eqnarray}

with $\Omega_M=0.25$ and $\Omega_{\Lambda}=0.75$, it reads

\begin{equation}
D_{\Lambda CDM}= \frac{c}{H_0}z (1+0.8125z-0.2422z^2)
\label{Lcdm1}
\end{equation}

To obtain the absolute magnitude from the measured apparent magnitude displayed in the data, we use
\begin{equation}
M=m-5 \log_{10}D_j -K -25
\label{mag}
\end{equation}

Here $D_j$ is in Mpc and $j$ denotes each of our three models: $j=E,EdS,\Lambda CDM$.

According to Surendran (2004), a reasonable approximation for the average K-correction applicable to the Sloan catalog is

\begin{equation}
K = -2.5(1+\alpha)\log_{10}(1+z),
\end{equation}

with $\alpha= - \, 0.5$. Therefore, we limit our sample to those galaxies that correspond to:

\begin{equation}
M_{lim}=17.77-25+1.25\log_{10}(1+z_{max})-5\log_{10}(r_{max}).
\end{equation}

Then, we examine the distribution function of the different input parameters. This fixes the range of the domain in which we perform our calculations. This limiting range, $d_{max}$, is the distance from the inner region to the outer region. We also determine, for each model, the maximum distance, $r_{max}$, from our galaxy to the most distant in our sample.

After having turned the RA, DEC, $z$ coordinates into Cartesian coordinates for each object and for each model, we make histograms of these coordinates, $X,Y,Z$, to obtain the final limits on our samples. These histograms yield approximate ranges. Combining this information with the number of inner and outer galaxies in a number of ranges, extracted from the domain representing our initial sample, and choosing the one with the highest number of galaxies, we obtain the best choice of the final sample to which we apply our analysis.

We start with \textbf {the Euclidean case}.

\begin{figure}
\centering
\includegraphics[height=6cm,width=8cm]{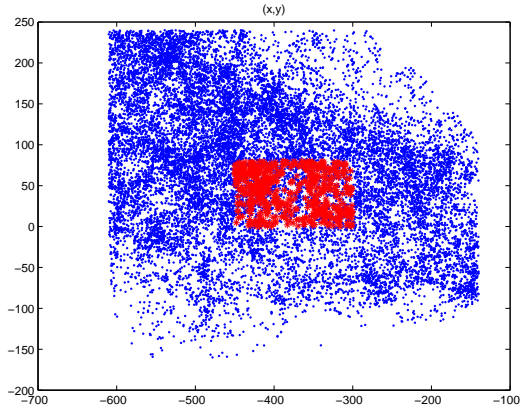}
\caption{The histogram of X,Y in the Euclidean case. In blue, the initial domain. In red, the final sample.}
\label{Fig2}
\end{figure}

As an example we give in Fig.~$\ref{Fig2}$, the $X,Y$ domain, for this model.

The histograms yield the following ranges (in Mpc):

 $$ -450<X<-300, \; \; 0<Y<80, \; \; -320<Z<-120,$$

and $r_{max}$=685.6 Mpc comes from Eq.~(\ref{euclid}) where the redshift $z$ is set to its maximum value, 0.16.

As a result our sample is reduced to 21,488 galaxies. We fix also, as explained in Sec.~\ref{cim}, an inner zone of 742 galaxies.

\begin{figure}
\centering
\includegraphics[height=6cm,width=8cm]{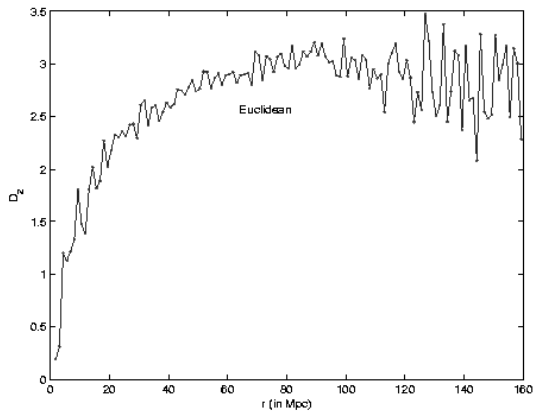}
\caption{The Euclidean case: $D_2 $ as a function of the scale $r$, obtained by least square fits for three points at a time.}
\label{Fig3}
\end{figure}

Then we proceed as follows. We divide the range $d_{max}$ into 256 segments. We apply the Grassberger-Procaccia method to each segment and obtain the dimension $D_2$ for each of them. We take three consecutive segment ends and perform on them a least square calculation of $D_2$. Then we take the last of these three and add two more points for the next least square calculation. Therefore we end up with 128 different dimensions. The final correlation dimensions are calculated by taking a consistent average $D_2$ on each range where this is feasible, which allows us to obtain constant dimensions on rather large scale ranges. By completing such an averaging procedure we are also able to calculate the root mean square deviation, and so to obtain a reasonable error measure. We find impossible to derive a robust constant dimension below 52 Mpc, but in the range 52 to 124 Mpc we obtain: $D_2=2.98 \pm 0.02$. Below 52 Mpc, $D_2$ increases from about 1 to 3. Above 124 Mpc, the influence of the edge effects starts to bring into play, so we have to wait for larger catalogs to go beyond this scale. The results for this Euclidean case are plotted in Fig.~$\ref{Fig3}$.

They are in a quite good agreement with that obtained by Hogg et al. (2005) for the SDSS luminous red galaxies.

Some authors (e. g., Tikhonov, 2006) used the $\Gamma$, spherical layers method, which was first introduced by Pietronero (1987). A discussion concerning this approach can be found in (Thieberger, Spiegel and Smith, 1990). We found no advantage to use this method instead of the correlation integral. Still, we thought interesting to check with it our results for various domain division choices. In Table 1 we show the $D_2$ values obtained for coarser and finer divisions. We see that going to a finer division does not improve the results. We conclude that this might be an additional confirmation that the integral method is good enough.

\begin{table}
\caption{$D_2$ values obtained for different numbers of domain divisions, $N_{BIN}$.}
\begin{tabular}{cccc}
\hline
$N_{BIN}$&$64$&$256$&$1024$\\
\hline
$D_2$&$2.95 \pm 0.02$&$2.98 \pm 0.02$&$2.9 \pm 0.2$\\
\hline
\end{tabular}
\end{table}

We now turn to \textbf{the Einstein-de Sitter model}. 

Here, the high density region remains roughly the same as in the Euclidean case. Therefore, for comparison sake, we choose the range:

 $$ -450<X<-300, \; \; 0<Y<80, \; \; -340<Z<-140,$$

and obtain a sample reduced to 19,027 galaxies, for which the inner region exhibits 629 galaxies. We obtain $r_{max}$=711.03 Mpc from Eq.~(\ref{eds}) where $z$ is set to its maximum value, 0.16. The average value of the correlation dimension in the range 52 to 124 Mpc (just as in the Euclidean model) comes out to be $D_2= 2.95 \pm 0.02$.

\begin{figure}
\centering
\includegraphics[height=6cm,width=8cm]{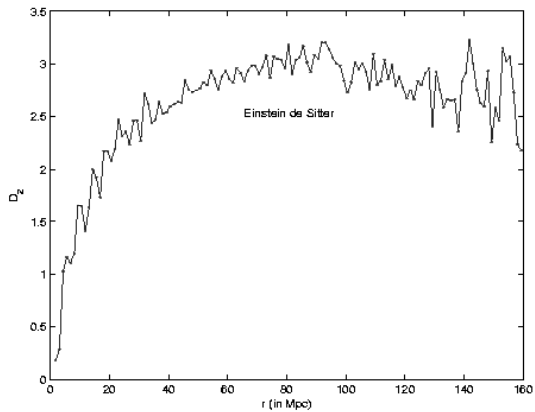}
\caption{The Einstein-de Sitter model: $D_2 $ as a function of the scale $r$, obtained by least square fits for three points at a time.}
\label{Fig4}
\end{figure}

We notice that the results are rather similar than for the previous cosmological model. They are given in Fig.$\ref{Fig4}$. It is worth stressing that the transition to homogeneity seems less complete in this case than in the Euclidean model (the fractal value remains somewhat under 3), but the difference is quite minor.

Now we consider the last case, \textbf{the $\Lambda$CDM model}. 

The preliminary results look less good. The edge effects start earlier and so at $d_{max}=160$ Mpc the error on $D_2$ becomes very large. Therefore, we limit ourselves to a smaller region, with $d_{max}$=120 Mpc. In this case the range taken into consideration is:

 $$ -400<X<-300, \; \; 0<Y<80, \; \; -280<Z<-130,$$

\begin{figure}
\centering
\includegraphics[height=6cm,width=8cm]{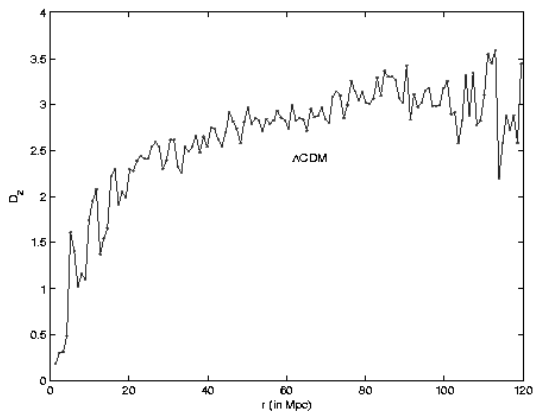}
\caption{The $\Lambda$CDM model: $D_2 $ as a function of the scale $r$, obtained by least square fits for three points at a time.}
\label{Fig5}
\end{figure}

Under these conditions we get again very similar results, but the errors
outside the chosen range are much larger. The sample is here reduced to 11,698 galaxies with 390 galaxies in the inner region. From Eq.~(\ref{Lcdm}), the value for $r_{max}$ reduces to 676.3 Mpc and between 45 Mpc and 104 Mpc we obtain: $D_2=2.97 \pm 0.02$.The analysis results are shown in Fig.$\ref{Fig5}$. We notice that the fluctuations are larger than in the previous cases, especially at small scales.

However, we can infer from the similarity of the results obtained with these three very different cosmological models that our method is rather model-independent. This strengthens the accuracy of our choice to neglect redshift-space distortions to perform this analysis.

\section{Check of the Method: Application to a Random Map Catalog} \label{armc}

To check the robustness of our method we apply it now to a random map.

We prepare a mock catalog where the ``galaxies'' are randomly distributed within the limits set by the real catalog. Since we have seen that our method is almost model-independent, we choose the EdS model as the prototype for our random map. We calculate the absolute magnitude, $M$, of each object from Eq.~(\ref{mag}) where the expression for the luminosity distance,, $D_j$, is given by Eq.~(\ref{eds}). This enables us to build a histogram of the $M$ values. We then fit the histogram to some function, which we use to obtain random $M$ values distributed just as the $M$'s pertaining to the EdS model. The redshift and angles determining the sample are chosen randomly:

\begin{equation}
RA=rdm_1(RA_{max}-RA_{min})+RA_{min}
\label{ra}
\end{equation}

\begin{equation}
DEC=rdm_2(DEC_{max}-DEC_{min})+DEC_{min}
\label{dec}
\end{equation}

\begin{equation}
z=C\sqrt{rdm_3^2+rdm_4^2+rdm_5^2}
\label{zz}
\end{equation}

Here, $rdm_i$ denotes random numbers distributed between 0 and 1, and $C$ is chosen so as to correspond to the distribution of the observed redshifts. We use a random number generator so that we obtain the same number of ``galaxies'' as in the volume limited catalog. We consider the same $d_{max}$ and use the same calculation method we apply to this mock catalog.

\begin{figure}
\centering
\includegraphics[height=6cm,width=8cm]{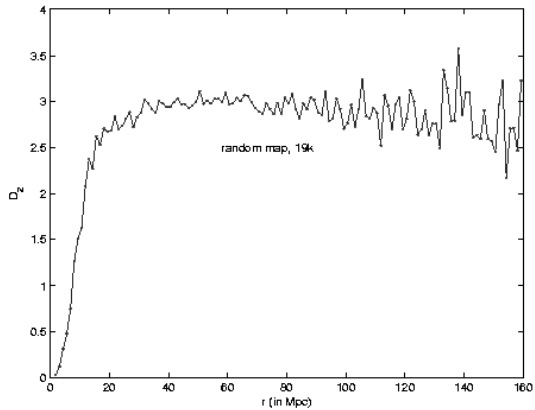}
\caption{The small random map: $D_2 $ as a function of the scale $r$, obtained by least square fits for three points at a time.}
\label{Fig6}
\end{figure}

The plot of the correlation dimension $D_2$ as a function of scales for this random map is displayed in Fig.$\ref{Fig6}$.

We see that in the small $r$ region this dimension is increasing faster than for the observed catalog analyzed in the framework of a similar EdS model. Actually, $D_2=3$ is reached around 35 Mpc. For the range 35 to 100 Mpc, we obtain $D_2= 2.96 \pm 0.01$.

\begin{figure}
\centering
\includegraphics[height=6cm,width=8cm]{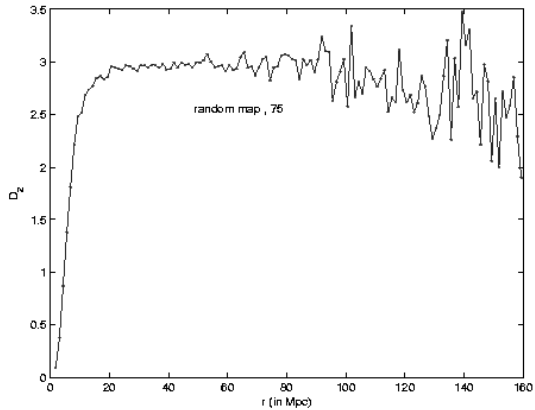}
\caption{The large random map: $D_2 $ as a function of the scale $r$, obtained by least square fits for three points at a time.}
\label{Fig7}
\end{figure}

Since the correlation dimension we should obtain at all scales for a randomly distributed point set in a 3-dimensional volume is $D_2=3$, we suspect this small scale anomaly might be due to the sparse population of these scales. Thus, we repeat the calculation for a point number which is four times as large.  The results are given in Fig.$\ref{Fig7}$.

Because we have more points, the dimension 3 is reached at lower scale.
At the large scale limit, the influence of edge effects implies there is no significant change. Here we obtain, for the range 20Mpc to 100Mpc, $D_2= 2.97 \pm 0.01$.

We therefore conclude that we must be very cautious about what we can infer from such an analysis at scales below 20-30 Mpc.

\section{Discussion and Conclusion} \label{dc}

We have used the publicly available data from the DR5 SDSS catalog to complete an analysis of the dimension $D_2$ of the galaxy distribution with the correlation integral method. Actually, there exists another method well designed to deal with the edge effect and described in (Coleman and Pietronero,1992). But although our method results in having a smaller value, $d_{max}$, for the larger probed scales, it seems safer as less approximations are involved (Provenzale et al., 1997).

To check the sensitivity of this method as regards redshift-space distortion or light cone effects, we performed our analysis in the framework of three different cosmological models. It came out that the dimension $D_2$ as a function of scales obtained in these different cases exhibits almost the same behavior, strengthening therefore our feeling that these effects could be safely ignored.

We performed another check of the robustness of the method by comparing the analysis of the observational catalog with that of random maps. We could thus see that at small scales, the random catalogs exhibit a fast increase of $D_2=$ toward the value 3 pertaining to a homogeneous distribution in a 3-dimensional volume, much steeper than the increase obtained with the real catalog. However, since $D_2=3$ is only reached around some 20-30 Mpc with the mock catalogs, we stress we must be cautious about what we can infer from such an analysis at scales below these values. Moreover, we can conclude from such results that, to obtain a more reliable description of the cosmological structures at various scales, it will be mandatory to use the larger samples which will be available in the future.

Meanwhile, the application of our method to the galaxy sets we extracted from the SDSS DR5 main sample yielded a correlation dimension $D_2$ increasing from very small values up to $D_2=3$ around 70-80 Mpc. This is compatible with the results obtained by Hogg et al. (2005) for luminous red galaxies and by Tikhonov (2006) for the DR4 SDSS Main Galaxy Sample.

In the smaller scale range where our results can be considered as rather robust, i. e., for 20 Mpc $<r<$ 70 Mpc, we noticed a slow increase from $D_2=2$ to $D_2=3$. This could be the signature of the galaxy distribution not being monofractal (or self similar) in this range, i. e., the correlation dimension is not a constant but varies with scale. A discussion of this issue, applied to the galaxy distribution, can be found in McCauley (2002). Such a property was expected from previous results obtained by, e. g., Murante et al. (1998).

It might also explain the discrepancies between the results of previous analyses performed with different samples probing different or too large scale ranges and for which self similarity was posed as an a priori assumption (Davis and Peebles, 1983; Coleman, Pietronero and Sanders, 1988; Sylos Labini, Montuori and Pietronero, 1998; Wu, Lahav and Rees, 1999) . Actually, if the correlation dimension varies with scales, the monofractal hypothesis implies that the results obtained by these various works must be different.

Now, one must be aware that what we have examined here is the distribution of bright matter only. In standard cosmology, one usually admits that most of the baryonic matter is present under the form of ``dark matter'', which, since it emits no light, cannot be seen in surveys such as the SDSS. We therefore claim that our results apply to bright matter only and that their extension to the whole baryonic matter content of the Universe would be unwise without further proof that ``light traces mass''. \\

{\it Acknowledgements}

We wish to thank Professors E.A. Spiegel and L. Nottale for valuable suggestions and discussions, and the anonymous referee for interesting comments and suggestions.

Funding for the creation and distribution of the SDSS Archive has been provided by the Alfred P. Sloan Foundation, the Participating Institutions, the National Aeronautics and Space Administration, the National Science Foundation, the U.S. Department of Energy, the Japanese Monbukagakusho, and the Max Planck Society.

The SDSS is managed by the Astrophysical Research Consortium for the participating institutions. The participating institutions are the University of Chicago, Fermilab, the Institute for Advanced Study, the Japan Participation Group, Johns Hopkins University, Los Alamos National Laboratory, the Max-Plank Institute for Astrophysics, New Mexico State University, University of Pittsburgh, Princeton University, the United States Naval Observatory, and the University of Washington.

\end{document}